# GateFormer: Forecast-Space Decomposition for Intra-Hour Solar Irradiance Prediction from Sky Images

David Hamlyn [a,*], Sunny Chaudhary [a], Jonathon Hare [a], Tasmiat Rahman [a]

[a] *School of Electronics and Computer Science, University of Southampton, University Road, Southampton, SO17 1BJ, United Kingdom*

## Abstract

Sub-hourly dispatch, balancing, and storage control increasingly depend on forecasts that anticipate rapid variability in solar generation. However, accurate sky-image-based irradiance forecasting requires a combination of visual cloud dynamics with recent irradiance state while adapting to different short-term irradiance-change regimes, a challenge for many existing approaches. To address this, we introduce GateFormer, a GHI-conditioned dual-decoder Vision Transformer (ViT) for intra-hour global horizontal irradiance (GHI) forecasting through forecast-space decomposition. GateFormer embeds observed GHI into the visual token stream, allowing interaction with sky-image patch features through self-attention. From this shared representation, a Trajectory Head predicts future GHI sequence, while a Delta Head predicts final-horizon change from the latest observed GHI. The two final-horizon forecasts are adaptively weighted in prediction space, with magnitude- and direction-supervision targeting the remaining error. On the NREL-TSI benchmark, GateFormer attains 125.01 W/m² RMSE at +30 min, improving over a state-of-the-art transformer baseline by 9.6% under paired-timestamp evaluation. Across NREL-TSI, Chilbolton, and SIRTA, GateFormer achieves the lowest final-horizon RMSE among the evaluated baselines, with forecast skill of 22.97–28.86% relative to smart persistence. Component ablations show adaptive weighting outperforms both individual decoders and fixed averaging, distinguishing GateFormer from conventional forecast combination, while the supervised correction further reduces RMSE by 1.68 W/m² and MAE by 4.86 W/m². Ramp-regime diagnostics show larger gains during operationally important positive ramps, reducing RMSE by 26.27 W/m². These results identify forecast-space decomposition as an alternative forecasting formulation, with potential relevance to other multimodal short-horizon forecasting tasks and deployment using only sky images and measured GHI.

## 1. Introduction

As photovoltaic (PV) penetration increases, power-system operation becomes more sensitive to rapid changes in solar generation. This sensitivity is becoming more operationally important as electricity systems move toward higher-granularity operation, including 5–15 min dispatch, sub-hourly balancing, and increasingly dynamic intraday markets [1],[2],[3],[4]. Sub-hourly irradiance forecasts are therefore important for grid balancing, storage scheduling, ramp-rate control, and short-term energy management [1],[2],[3],[4]. At these horizons, forecast errors often arise from local cloud advection, transient solar obstruction, and cloud-edge enhancement, which can rapidly change ground-level irradiance. Conventional numerical weather prediction (NWP) often falls short at this resolution due to spatial and temporal scales below those of typical operational weather-model outputs [5]. Ground-based sky images provide direct observations of the local cloud field, including cloud structure, motion cues, solar occlusion, and cloud-edge patterns, making them a valuable input for intra-hour global horizontal irradiance (GHI) forecasting [4],[6],[7],[8]. Accurate intra-hour forecasting therefore requires visual cloud information to be interpreted together with the current irradiance state.

Recent deep learning (DL) approaches have improved intra-hour irradiance forecasting by learning predictive representations directly from sky-image sequences [9], [10]. These advances include future visual-state prediction, transformer-based sequence modelling, and multimodal fusion of sky imagery with irradiance history and clear-sky information [9],[10],[11],[12]. However, strong performance in this setting still depends on how the model represents short-term irradiance change, particularly when cloud motion produces rapid departures from persistence-like behaviour. This motivates a forecasting formulation that does not rely on a single output pathway but instead forms distinct predictions of future irradiance state and learns how to effectively combine them.

∗ Corresponding author. Email: dh13g23@soton.ac.uk

To address this limitation, we introduce GateFormer, a dual-decoder Vision Transformer (ViT) that implements adaptive gating in the prediction space for intra-hour irradiance forecasting. Rather than using a single decoder to map encoded sky-image information directly to the final forecast, GateFormer first encodes the joint visual and irradiance state by embedding observed GHI into the visual token stream, allowing irradiance-state information to interact with sky-image patch features through self-attention. From this shared representation, a Trajectory decoder predicts GHI at each future step up to and including the final horizon, meaning the furthest point in time forecast by the model, while a Delta decoder predicts the change from the latest observed GHI directly to that same final horizon. The forecast-space decomposition means forming alternative forecasts with different prediction targets from a shared representation and exposing their values and disagreement to the gate before the final prediction is formed. The two final-horizon predictions allow a shared architecture to model complementary views of the prediction space which are then combined using a learned adaptive blend to form a Base forecast, and a residual head predicts the remaining post-blend correction using magnitude and direction supervision. Unlike conventional ensembles that combine independently trained models, or Mixture-of-experts (MoE) architectures that route inputs among alternative expert pathways, GateFormer forms both candidate forecasts from one shared representation using distinct prediction targets and applies gating only after explicit irradiance predictions have been produced. This structure allows the model to represent both gradual trajectory evolution and direct final-horizon displacement, adaptively blend them in prediction space and improve post-blend error.

**Contributions and motivations**. Reliable intra-hour irradiance forecasting is critical for ramp-rate control, storage scheduling, and grid stability, yet multimodal methods for forming and combining complementary predictions remain comparatively underexplored. Building on state-of-the-art ViTs, we introduce GateFormer, a novel forecast-space decomposition architecture. (1) A dual-decoder framework predicts absolute multi-horizon GHI trajectories and direct final-horizon GHI displacement from a shared GHI-conditioned visual representation, allowing the model to capture both stable trajectory behaviour and larger ramp displacements. (2) A prediction-space adaptive blending that learns a per sample combination of the two forecasts, rather than fusing latent features, benefiting from the decoders' complementary representations. (3) A separately supervised post-blend residual that predicts the magnitude and direction of the error remaining after adaptive gating. Experiments across three sky-imaging datasets and component-level ablations demonstrate the learned blend improves over both individual decoders and fixed averaging, while residual correction provides a further overall reduction in error especially in positive ramps.

## 2. Related Work

Short-term solar forecasting has increasingly used ground-based sky images as they provide local information on cloud position, motion, and solar obstruction that is unavailable from scalar irradiance histories alone [4]. Classical methods convert these observations into short-term forecasts through cloud segmentation, optical flow, and tracking [6],[7]. Although physically interpretable, their performance depends on preprocessing quality and assumptions concerning cloud motion, height, and transmittance, motivating DL approaches that learn predictive visual representations directly from imagery [4], [5], [6], [7].

Recent studies have applied explainable machine learning to GHI estimation [13], wavelet-assisted deep learning to GHI modelling [14], and feature-engineered machine learning to rainfall prediction [15]. Solar system modelling has also been explored through Arduino-based PV water-pumping simulations [16].

Self-attention enables long-range relationships to be modelled across input sequences [17], while ViTs extend this mechanism to spatial relationships between image patches [18]. Image-only ViTs have shown that sky-image representations can support accurate irradiance estimation [19]. The Solar Irradiance Anticipative Transformer (SIAT) extends this capability to temporal forecasting from sky-image sequences [10]. Paletta et al. jointly predict future visual states and irradiance to anticipate cloud-induced perturbations [9]. However, visual information alone does not explicitly encode current irradiance state or its recent evolution. Piantadosi et al. demonstrate the value of non-image inputs for Transformer-based PV forecasting [20]. Liu et al. further demonstrate this by fusing sky-image features with historical and clear-sky GHI through cross-modal attention, reporting an nRMSE of 4.28% at the 10-min horizon on NREL SRRL [11]. Zhang et al. compare early, late, and gated image–numerical fusion, demonstrating trade-offs between overall error and ramp-event performance. Their Transformer-based early fusion improves ramp-event balanced accuracy by 9.43% and 3.91% at the 2- and 6-min horizons, respectively [12]. These approaches improve representation and modality interaction but still optimise a single forecast mapping rather than forming complementary predictions with different forecasting objectives.

Architectures have begun to separate modelling roles through specialised pathways, including sequential and parallel frameworks to extract distinct visual and irradiance information [21]. Specialisation also appears in the handling of heterogeneous inputs, with Chen et al. combining spatiotemporal satellite and NWP inputs within an interpretable Temporal Fusion Transformer for intra-day irradiance forecasting [22]. At the processing-stage level, neural feature extraction has been separated from probabilistic prediction [23]. More explicit branching appears in PV-MM-Diffusion, which couples image and PV pathways to jointly generate future sky states and probabilistic PV trajectories [24], while Xiong et al. generate multiple future sky scenarios, forecast power for each, and aggregate the resulting distributions [25]. These studies demonstrate the value of architectural specialisation, but their components separate modalities, processing stages, or deterministic and probabilistic dynamics; they do not generate alternative irradiance forecasts from the same latent representation and combine them in prediction space. This distinction is technically important because retaining both candidates as internal forecasts exposes their values and disagreement to the gate before the final prediction is formed, unlike latent fusion that produces one output or conventional combination applied to completed forecasts from separate systems.

These specialised pathways can be combined adaptively, with dual gates improving interactions among spatial and temporal features [26]. Forecasts from wavelet-decomposed PV subseries have also been weighted dynamically across multiple horizons [27]. MoE models formalise input-dependent weighting of specialised predictors [28], while forecast-combination theory establishes that combining imperfect forecasts outperform individual predictors [29]. Nouri et al. apply this principle to solar forecasting blending separately generated ASI, satellite, and persistence forecasts, reducing RMSE and MAE by 13.6% and 17.0%, respectively, relative to standalone satellite forecast [30]. This demonstrates the value of forecast-level complementarity, although candidate forecasts originate from separate forecasting systems.

Collectively, these studies establish the value of multimodal context, specialised pathways, gating, and forecast-level combination. What remains unresolved is whether a shared visual–irradiance representation can support multiple forecast objectives, allow their final-horizon predictions to be weighted conditionally on the input, and explicitly correct the error remaining after combination. GateFormer addresses this gap by coupling a shared GHI-conditioned visual backbone with a Trajectory Head for multi-horizon evolution and a Delta Head for final-horizon displacement, followed by adaptive forecast-space blending and supervised residual correction. Unlike approaches requiring additional measured meteorological variables or satellite products, GateFormer uses sky-image sequences and observed GHI, supplemented only by deterministically derived temporal and clear-sky context.

## 3. Method

We introduce GateFormer by first outlining the overall architecture in 3.1, then describing the dual forecasting heads in 3.2, the adaptive blend-residual mechanism in 3.3, the loss functions in 3.4, and the datasets and evaluation protocol in the following sections.

### *3.1. Overall Architecture*

Figure 1 shows the GateFormer pipeline, $\mathcal{Z}_{\mathrm{img}}$ denotes the encoded sky-image sequence representation, $\mathcal{Z}_{\mathrm{phys}}$ denotes the temporal and clear-sky context, $\omega$ is the adaptive blend weight, and $\hat{\mathrm{r}}_{\mathrm{t+H}}$ is the predicted post-blend residual correction. Let t denote the latest observed time, $h \in \{1, \dots, H\}$ the forecast step, and H the final forecast step. In this study, H=3. Each input sample consists of four past sky images, their corresponding GHI anchor values, and temporal/clear-sky features. A DeiT-III encoder maps each image into patch tokens [31]. For each input frame the observed GHI value is projected into the transformer embedding space and inserted as a learned GHI token, allowing irradiance state information to interact with image patch representations through self-attention.

Frame-level time encodings are added to the image-token representations, while clear-sky and recent-GHI-change features condition the decoders and adaptive gate/residual module. The resulting patch-level memory is pooled into frame-level representations, and temporal-difference tokens are formed from consecutive frame summaries and the latest-minus-first frame summary before being passed to two lightweight decoders: a multi-step Trajectory decoder and a Delta displacement decoder. The decoder memory contains both per-frame visual state information and short-term visual-change information across the four-image sequence. Their final-horizon predictions are adaptively blended, and a residual head predicts the remaining post-blend correction.

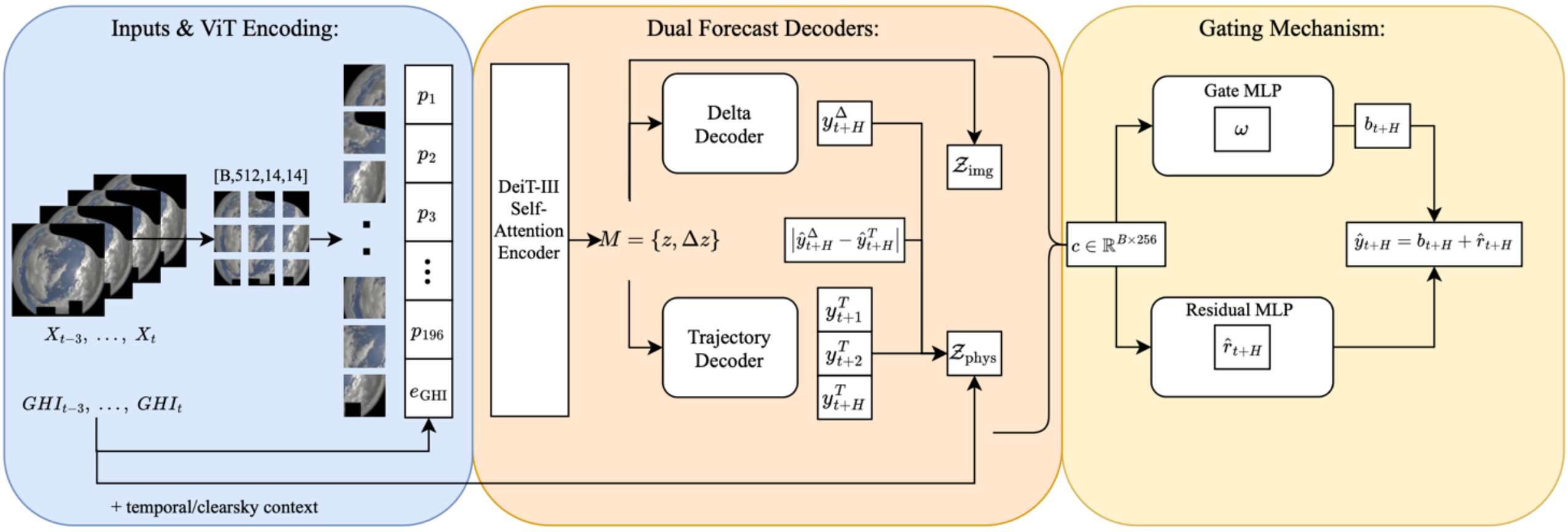

Figure 1 – Overview of the GateFormer architecture. A GHI-token-conditioned ViT encodes the sky-image sequence and irradiance history. A Trajectory Head and Delta Head produce complementary forecasts, which are combined through an adaptive blend and refined by a residual correction.

Table 1 – Architecture-summary table

| Component | Input | Configuration | Output | Param (M⁺) |
|---|---|---|---|---|
| Shared Encoder | Four 224 by 224 images corresponding GHI | DeiT-III Medium: 16 by 16 patches, d=512, 12 Blocks, 8 Heads, MLP=2048 | $M \in \mathbb{R}^{B\times 8\times 512}$ | 38,347,804 |
| Trajectory Head | Decoder memory M, learned query qT | Two-layer Transformer decoder: d=512, 8 Heads, MLP=2048, dropout 0.1 | $[\hat{y}^{T}_{t+1}, \hat{y}^{T}_{t+2}, \hat{y}^{T}_{t+H}]$ | 8,410,627 |
| Delta Head | Decoder memory M, learned query qD | Two-layer Transformer decoder (dim=Trajectory Head) | $\hat{y}^{\Delta}_{t+H}$ | 8,409,601 |
| Multimodal Context | 18 numeric features, three visual summaries | Numeric 18 → 256, visual 1536→256, fused 512→256; LayerNorm, SiLU, dropout 0.05 | $c \in \mathbb{R}^{B\times 256}$ | 533,796 |
| Adaptive Gate | Context vector $c$ | Linear 256→1 followed by sigmoid | $\omega \in [0,1]$ | 257 |
| Residual Head | Context vector $c$ | Separate Linear 256→1 unbounded | $\hat{r}_{t+H}$ | 257 |
| GateFormer | All components above | Shared encoder, two forecast heads, adaptive gate and residual correction | $\hat{y}_{t+H}$ | 55,702,342 |

For input frame k, the encoder output is pooled into a frame-level representation $z_k$, and the four frame representations are combined with consecutive and full-window temporal differences to form the shared decoder memory M:

$$z_k = \text{Pool}\big[E_\theta\big([P_k; e_G(g_k)] + e_t(k)\big)\big] \tag{1}$$

$$\text{M} = [z_1, z_2, z_3, z_4, z_2 - z_1, z_3 - z_2, z_4 - z_3, z_4 - z_1] \tag{2}$$

Here, $P_k$ denotes the patch tokens, $e_G(g_k)$ the GHI token, $e_t(k)$ the temporal encoding, $E_\theta$ the encoder, and Pool token averaging. The resulting memory M contains four frame-state representations, three consecutive-change representations, and one full-window change representation. This shared memory is supplied to both forecasting heads.

### *3.2. Dual Heads*

The shared visual-temporal encoder feeds two complementary forecasting heads. The first is the Trajectory Head, which predicts the full future irradiance sequence across all forecast steps. This head is designed to capture the temporal evolution of GHI over the forecast window, including gradual trends and horizon-dependent changes in irradiance. Here, gradual trends refer to progressive variation in GHI across successive forecast steps, while horizon-dependent changes mean that each lead time has its own predicted GHI value. The second is the Delta Head, which focuses directly on the final forecast step. Rather than predicting the entire future path, it predicts the final horizon change relative to the latest observed GHI. Here, the final-horizon change is the signed difference between GHI at the final forecast step and the latest observed GHI, and this predicted difference is added to the latest observation to produce the Delta forecast. This gives the model an alternative final-horizon decoder that directly models displacement from the latest irradiance state, rather than deriving the final forecast from the multi-horizon Trajectory output. Together, these two heads provide complementary views of the forecasting problem. The adaptive blend is therefore the central decision mechanism: it determines, for each sample, whether the final forecast will rely more on the trajectory-derived forecast or the direct displacement forecast.

$$\text{u}^T = \text{Dec}_T(q_T, \text{M}) \tag{3}$$

$$\text{u}^\Delta = \text{Dec}_\Delta(q_\Delta, \text{M}) \tag{4}$$

Output projections map $u^T$ to the multi-horizon Trajectory forecast and $u^\Delta$ to the predicted final-horizon change $\hat{y}^\Delta_{t+H}$. Let $\Delta y_{t+H} = y_{t+H} - y_t$ denote the measured final horizon change and $\Delta\hat{y}_{t+H}$ its prediction. The corresponding absolute GHI forecast is $\hat{y}^\Delta_{t+H} = y_t + \Delta\hat{y}_{t+H}$.

### 3.3. Soft Gating Mechanism

In GateFormer, the gating module consists of an adaptive prediction-space blend followed by a residual correction stage. The blend estimates the relative weighting of the Trajectory and Delta forecasts, while the residual head predicts the remaining post-blend error. The Trajectory Head and Delta Head produce two complementary predictions for forecast horizon H. Let $\hat{y}^T_{t+H}$ denote the Trajectory Head prediction and $\hat{y}^\Delta_{t+H}$ denote the Delta Head prediction. The adaptive blend weight is predicted from a context vector containing the encoded sky-image sequence representation, recent GHI-state information, the Trajectory Head forecast, the Delta-derived final-horizon forecast, and the disagreement between the two decoder outputs. Here, $\mathcal{Z}_{\mathrm{img}}$ contains the three visual summaries, while $\mathcal{Z}_{\mathrm{phys}}$ contains the numerical, clear-sky, and decoder-derived forecast features. These are fused as:

$$c = MLP_{fuse}([\mathcal{Z}_{\mathrm{img}}, \mathcal{Z}_{phys}]) \tag{5}$$

The gate therefore does not operate directly on raw images or raw irradiance values; instead, it uses the learned visual-temporal representation together with the two-candidate final-horizon predictions to determine their relative contribution to the Base forecast. Rather than selecting one decoder output with a hard decision, GateFormer learns a sample-specific blend weight $\omega \in (0,1)$ which forms the Base forecast:

$$b_{t+H} = \omega\hat{y}^T_{t+H} + (1-\omega)\,\hat{y}^\Delta_{t+H}, \qquad \omega \in [0,1] \tag{6}$$

The blend weight $\omega$ determines the relative contribution of the trajectory-derived and delta-derived predictions for each sample. A value close to 1 places greater weight on the Trajectory Head, while a value close to 0 places greater weight on the Delta Head. This allows the model to combine the two decoder outputs in prediction space after both forecasts have already been formed. After the base forecast is produced, a residual head predicts the remaining post-blend error. Let $\hat{r}_{t+H}$ denote the predicted residual correction. The final corrected forecast is then

$$\hat{y}_{t+H} = b_{t+H} + \hat{r}_{t+H} \tag{7}$$

This design separates forecast formation into two stages. The adaptive blend first combines the two decoder predictions into the Base forecast, and the residual then corrects the remaining error after blending. The blend is not directly supervised to select the locally better decoder. Instead, the blend weight ($\omega$) and residual correction are learned through the final prediction objective and the residual-supervision terms described below.

### 3.4. Loss Functions

Training follows two sequential stages with separate objectives. First, the shared encoder, Trajectory Head, and Delta Head are trained using the decoder-pretraining objective. These components are then frozen, and the adaptive blend-residual module is trained using the gate objective. The Trajectory loss and the change- and event-weighting terms are defined as follows:

$$\mathcal{L}_{\mathrm{Traj}} = \sqrt{\mathbb{E}_h\left[\left(y_{t+h} - \hat{y}^T{}_{t+h}\right)^2\right] + \varepsilon} \tag{8}$$

$$\mathcal{W}^{chg} = 1 + \lambda_{chg}\mathrm{clip}\left(\frac{|\Delta y_{t+H}|}{s_{chg}}, 0, c_{chg}\right) \tag{9}$$

$$\mathcal{W}^{event} = (|\Delta y_{t+H}| \geq \tau_{event})\mathrm{clip}\left(\frac{|\Delta y_{t+H}|}{s_{event}}, 0, c_{event}\right) \tag{10}$$

Before evaluating the losses, $\mathcal{W}^{\mathrm{chg}}$ and $\mathcal{W}^{\mathrm{event}}$ are divided by their respective mini-batch means. Their normalised forms are denoted by $\widetilde{\mathcal{W}}^{\mathrm{chg}}$ and $\widetilde{\mathcal{W}}^{\mathrm{event}}$. Here, $\ell_{\mathrm{event}}$ denotes the positive underprediction error for large upward changes and the positive overprediction error for large downward changes. If a mini batch contains no qualifying event, $\mathcal{L}_{\mathrm{event}}$ is set to zero.

$$\mathcal{L}_{\Delta,base} = \sqrt{\mathbb{E}\left[\widetilde{\mathcal{W}}^{chg}(\Delta y_{t+H} - \Delta\hat{y}_{t+H})^2\right] + \varepsilon} \tag{11}$$

$$\mathcal{L}_{event} = \sqrt{\mathbb{E}[\widetilde{\mathcal{W}}^{event} \ell_{event}^2] + \varepsilon} \tag{12}$$

$$\mathcal{L}_{\Delta} = \mathcal{L}_{\Delta,base} + \lambda_{event}\mathcal{L}_{event} \tag{13}$$

$$\mathcal{L}_{Heads} = \mathcal{L}_{\mathrm{Traj}} + \mathcal{L}_{\Delta} \tag{14}$$

After decoder pretraining, the encoder and both forecasting heads are frozen. After the Base forecast $\mathrm{b}_{\mathrm{t+H}}$ is formed, the true post-blend residual is defined as $\mathrm{r}_{\mathrm{t+H}} = (\mathrm{y}_{\mathrm{t+H}} - \mathrm{b}_{\mathrm{t+H}})$. The blend-residual module is then trained using a final prediction loss, a residual-error loss, and a residual-direction loss:

$$\mathcal{L}_{\mathrm{fit}} = \sqrt{\mathbb{E}[(y_{t+H} - \hat{y}_{t+H})^2] + \varepsilon} \tag{15}$$

$$\mathcal{L}_{\mathrm{res}} = \sqrt{\mathbb{E}[(r_{t+H} - \hat{r}_{t+H})^2] + \varepsilon} \tag{16}$$

$$\mathcal{L}_{\mathrm{dir}} = \mathbb{E}[\mathrm{softplus}(-\mathrm{sgn}(r_{t+H})\, \hat{r}_{t+H})] \tag{17}$$

$$\mathcal{L} = \mathcal{L}_{\mathrm{fit}} + \mathcal{L}_{\mathrm{res}} + \mathcal{L}_{\mathrm{dir}} \tag{18}$$

The fit loss trains the final corrected forecast, the residual loss encourages the residual head to predict the remaining post-blend error using a stop-gradient residual target, and the sign loss penalizes corrections that move in the wrong direction. The total objective therefore trains GateFormer to form an adaptive Base forecast and then apply a directionally consistent residual correction. The equations above define the core objective used to train the adaptive blend and residual correction. Stage-specific weighting, event weighting, and decoder-specific loss terms used during optimization are described in the Implementation section.

## 4. Dataset

We evaluate GateFormer on three sky-imaging datasets with co-located GHI measurements: National Renewable Energy Laboratory Total Sky Imager (NREL-TSI), Chilbolton, and SIRTA. NREL-TSI provides hemispheric total-sky imagery from Golden, Colorado, USA (39.742 °N, 105.18 °W) at 10-minute cadence [32]. Chilbolton provides all-sky imagery from the CFARR facility in Hampshire, UK (51.145 °N, 1.437 °W) at 5-minute cadence [33], [34], [35]. SIRTA provides sky imagery and irradiance measurements from Palaiseau, France (48.713 °N, 2.208 °E) at 2-minute cadence [36]. All streams are time-aligned to paired image-radiometry timestamps, filtered for missing frames, and evaluated in GHI units of W/m², using data from 2017 to 2019. Sky images are resized to 224 by 224 pixels and standardised via ImageNet preprocessing [37]. In this study we adopt a deliberately minimalist augmentation scheme: aside from ImageNet normalisation, no rotation, colour perturbations, blurring, or similar transforms are applied. This keeps architectural comparison transparent. Investigating richer, sky-specific augmentations remains valuable future work. Sequences are constructed at the site's native cadence using four past frames and three future horizons, providing intermediate trajectory supervision while retaining a common final forecast step for the Delta Head and adaptive blend. After timestamp alignment, daylight filtering, split assignment, and sequence construction, the final datasets contained 68,740 NREL-TSI sequences, 158,461 Chilbolton sequences, and 365,811 SIRTA sequences. For GateFormer, the train/validation/test sample counts were 46,229/11,323/11,188 (NREL +30 min), 107,198/26,327/24,936 (Chilbolton +15 min), and 248,110/59,638/58,063 (SIRTA +6 min). We use a ±10s timestamp tolerance for pairing images with radiometry and drop sequences with any missing frame. Unless stated otherwise, metrics are reported on the last horizon (e.g., +30 min at NREL-TSI). This split was done into a training set and an evaluation dataset as well as a separate testing dataset. Days 5–9 of each month were used as the fixed test set, days 15–19 as the validation set, and the remaining eligible days as the training set. This intra-month partition follows the SIAT evaluation protocol [10], enabling direct comparison while ensuring that each subset spans all months and seasons. The split is applied before sequence construction; each split is generated independently, and no training sequences share images or GHI target timestamps with validation or test sets. A daylight evaluation filter retains sequences only when future target GHI values exceed 10 W/m² and days with maximum GHI below 50 W/m² are excluded.

## 5. Implementation and Evaluation Metrics

We report MAE, RMSE, and FS vs a reference forecaster using paired-timestamp evaluation. We use $\mathrm{FS}_{\mathrm{smart}}$ (clear-sky scaled persistence). For a reference forecaster $\hat{y}_{\mathrm{ref}}$, FS follows the RMSE-based meteorological definition relative to a reference forecaster [38],[39]. Clear-sky irradiance $\mathrm{G}^{\mathrm{cs}}$ is computed via pvlib [40] using the simplified Solis model [41], [42]. Simplified Solis is used only to construct the common clear-sky-scaled persistence reference and does not affect the GateFormer predictions themselves.

$$\mathrm{FS_{ref}}(\%) = 100\left(1 - \frac{\mathrm{RMSE}(\hat{y}, \mathrm{y})}{\mathrm{RMSE}(\hat{y}_{\mathrm{ref}}, \mathrm{y})}\right) \tag{19}$$

$$\hat{y}_{\mathrm{smart}}(t + \Delta) = y(t)\frac{G^{\mathrm{cs}}(t + \Delta)}{\max(G^{\mathrm{cs}}(t), \varepsilon)} \tag{20}$$

The backbone model employs a ViT pretrained on ImageNet [43]. Images are split into $16 \times 16$ patches, then passed through multi-head self-attention layers. We use the DeiT-III Medium backbone with two forecast decoders and a compact gating MLP [31]. Optimisation employs AdamW with a learning rate of $2 \times 10^{-4}$, weight decay $2.5 \times 10^{-4}$ batch size 8 [44]. Training follows a two-stage schedule. First, the shared encoder, Trajectory decoder, and Delta decoder are pretrained for 4 epochs with the backbone frozen except for the final 8 transformer blocks. The Trajectory decoder is supervised on the full future GHI sequence using RMSE loss in absolute irradiance space. Absolute GHI is retained as the prediction target to preserve a direct mapping from measured irradiance to future measured irradiance, consistent with the benchmarks used. Clear-sky information is provided as contextual input. In this study $\varepsilon = 10^{-8}$. To emphasise larger irradiance-change events, the Delta loss uses change-magnitude weighting of 2, scale 200 W/m², and cap 3. An additional event-asymmetric term ($\lambda_{event}$) with weight 0.15 penalises under-prediction during large upward changes and over-prediction during large downward changes, using a 150 W/m² event threshold, scale 250 W/m², and cap 4. These values were fixed training hyperparameters and may be further optimised for each site. Second, the encoder and both heads are frozen, and the adaptive blend-residual module is trained for 6 epochs using the blend-residual objective. Training is performed in CUDA automatic mixed precision on a single NVIDIA RTX 3090 GPU with 24 GB memory. All site splits, configurations, and random seeds are to be released in the public repository for exact reproducibility upon acceptance.

Table 2 – Core Training Hyper Parameter Table

| Setting | Value |
|---|---|
| Backbone | DeiT-III Medium |
| Parameters | 55.70 M |
| Batch size | 8 |
| Optimiser | AdamW |
| Learning rate | $2 \times 10^{-4}$ |
| Weight decay | $2.5 \times 10^{-4}$ |
| Head pretraining | 4 epochs |
| Blend-residual training | 6 epochs |

Hyperparameters were selected through preliminary experiments using validation RMSE on NREL-TSI. The resulting GateFormer architecture, optimisation settings and loss coefficients were then fixed across NREL-TSI, Chilbolton and SIRTA without site-specific tuning. The sensitivity analyses evaluate the resilience of the principal architectural and loss choices around these selected values.

# 6. Results and Discussion

We first evaluate GateFormer at three geographically distinct sites as described in the Dataset section. The same preprocessing pipeline, model configuration, and training protocol are used for each site, allowing performance to be compared across different camera optics, atmospheric conditions, and cloud morphology without introducing site-specific methodological changes. We reproduce the SIAT baseline using the official implementation provided by the authors. DeiT and ResNet-50 use the latest observed image, while ConvLSTM uses the last four. All evaluations are performed at each site's native sampling cadence, Step 1 and Step 3 represent +10 and +30 min for NREL-TSI, +5 and +15 min for Chilbolton, and +2 and +6 min for SIRTA.

Table 3 – Performance comparison across sites under paired-timestamp evaluation. MAE/RMSE reported in W/m²; FS reported in %.

| Future Timesteps | | Step 1 | | | Step 3 | | |
|---|---|---|---|---|---|---|---|
| Dataset | Model | MAE | RMSE | FS | MAE | RMSE | FS |
| NREL-TSI | Reference | 57.82 | 128.83 | – | 81.33 | 162.29 | – |
| NREL-TSI | DeiT | 78.98 | 119.85 | 6.97 | 89.33 | 140.17 | 13.63 |
| NREL-TSI | ConvLSTM | 97.94 | 143.54 | –11.42 | 126.62 | 175.88 | –8.37 |
| NREL-TSI | ResNet-50 | 73.39 | 114.07 | 11.46 | 82.38 | 134.29 | 17.25 |
| NREL-TSI | SIAT | 65.26 | 113.03 | 12.26 | 79.76 | 138.35 | 14.75 |
| NREL-TSI | GateFormer | 51.97 | 103.98 | 19.29 | 70.44 | 125.01 | 22.97 |
| Chilbolton | Reference | 55.93 | 126.41 | – | 76.14 | 150.69 | – |
| Chilbolton | SIAT | 70.57 | 107.54 | 14.93 | 72.66 | 114.60 | 23.95 |
| Chilbolton | GateFormer | 46.89 | 98.67 | 21.94 | 59.63 | 111.98 | 25.69 |
| SIRTA | Reference | 38.13 | 91.22 | – | 59.21 | 122.47 | – |
| SIRTA | SIAT | 41.19 | 72.86 | 20.13 | 50.05 | 88.20 | 27.98 |
| SIRTA | GateFormer | 28.98 | 71.70 | 21.40 | 40.92 | 87.13 | 28.86 |

Table 3 compares GateFormer with standard vision and sequence baselines. On NREL-TSI, GateFormer achieves the lowest RMSE at both forecast steps and improves the Step 3 RMSE from 138.35 W/m² for SIAT to 125.01 W/m². Final-horizon improvements are observed at each site, where GateFormer reduces SIAT RMSE by up to 9.64%. Across all three sites, GateFormer achieves the highest final-horizon FS, ranging from 22.97% to 28.86%, indicating that the proposed dual-decoder blend-residual structure improves performance across multiple sky-imaging datasets.

## 6.1. *Seasonal Performance*

To determine GateFormer's effectiveness across seasonal regimes, we split the NREL-TSI +30 min test predictions by meteorological season. To evaluate the framework under different irradiance ranges, sun-path conditions, and cloud morphologies, and to determine whether the residual correction provides an additional benefit beyond the adaptive Base forecast across the seasons as seen in Table 4.

Table 4 – Performance by season (Step = 3). RMSE in W/m²; $FS_{res}$ is forecasting skill rel. to Base blend (higher is better).

| Season | Delta Head | Trajectory Head | Base RMSE | Final RMSE | ΔRMSE | ΔMAE | $\Delta FS_{res}$(%) |
|---|---|---|---|---|---|---|---|
| S1 DJF (Dec – Feb) | 82.34 | 72.19 | 70.21 | 68.47 | −1.73 | −1.84 | +2.47 |
| S2 MAM (Mar – May) | 135.19 | 126.97 | 125.44 | 124.61 | −0.83 | −2.48 | +0.66 |
| S3 JJA (Jun – Aug) | 181.95 | 167.25 | 163.64 | 161.42 | −2.22 | −7.64 | +1.35 |
| S4 SON (Sep – Nov) | 112.43 | 107.58 | 105.15 | 103.14 | −2.01 | −6.34 | +1.91 |

Despite the absolute RMSE changes by season, the FS of both the Base Forecast and Final Forecast remains similar across each season, the range of FS for Base and GateFormer changes 2–3%. Demonstrating GateFormer performs reliably across the year, indicating resilience under sun-path and cloud-regime shifts. The residual gives a small but consistent extra lift on top of an already strong Base Forecast. The largest absolute seasonal RMSE occurs in JJA at 161.42 W/m², where the residual also provides the largest RMSE and MAE gains, improving RMSE by −2.22 W/m² and MAE by −7.64 W/m². The smallest RMSE gain occurs in MAM at −0.83 W/m², but the residual contribution remains positive in every season. Across seasons, the same component ordering is preserved: the Trajectory Head improves over persistence, the adaptive blend improves over the individual heads, and the residual final prediction provides the best seasonal RMSE. This indicates that the proposed decomposition remains useful under different seasonal irradiance ranges and cloud regimes.

## 6.2. *Ablations and Analysis*

Table 5–6 reports the component-level ablation on NREL-TSI at the final forecast step. This is done to isolate each part of GateFormer and its contribution to the final prediction. The "No GHI Token" variant removes token-level irradiance conditioning from the transformer encoder. The No GHI Token variant removes GHI only from the transformer token sequence. Historical GHI-derived information remains available through the decoder and gating context, so the ablation isolates token-level irradiance conditioning rather than removing all GHI information. "No Residual (Base)" keeps the adaptive blend but removes the final residual. "GatePersist" replaces the learned Delta Head with a persistence-based branch to evaluate whether the second pathway needs to be learned. "Equal-Average" replaces the learned blend with a fixed average of the Trajectory and Delta predictions. The individual Trajectory Head and Delta Head rows report the performance of each decoder alone, while the DeiT row provides the backbone-only baseline. The GHI-only variant receives the same four historical GHI observations and corresponding temporal/clear-sky context as GateFormer but receives no sky images. ΔRMSE and ΔMAE are reported relative to the full GateFormer framework, positive values indicate worse performance than the full model.

Table 5 – Ablation study on NREL-TSI (+30 min). ΔRMSE and ΔMAE are reported as error increases rel. to the GateFormer.

| Variant | RMSE | MAE | ΔRMSE | ΔMAE |
|---|---|---|---|---|
| GateFormer (Final) | 125.01 | 70.44 | – | – |
| No GHI Token | 125.44 | 71.09 | +0.43 | +0.65 |
| No Residual (Base) | 126.69 | 75.30 | +1.68 | +4.86 |
| GatePersist | 128.00 | 72.76 | +2.99 | +2.32 |
| Equal-Average | 129.62 | 77.42 | +4.61 | +6.98 |
| Trajectory Head | 129.23 | 79.12 | +4.22 | +8.68 |
| Delta Head | 139.36 | 84.81 | +14.35 | +14.37 |
| Plain DeiT | 140.17 | 89.33 | +15.16 | +18.89 |
| GHI-only | 143.37 | 86.62 | +18.36 | +16.18 |

The ablation confirms the role of each component in the proposed architecture. The Trajectory Head is the stronger individual decoder overall, reaching 129.23 W/m² RMSE compared with 139.36 W/m² for the Delta Head. However, the Delta Head is not redundant: it outperforms the Trajectory Head in the most extreme absolute-change regime, reducing RMSE from 319.48 to 294.74 W/m² for | ΔGHI | >250 W/m², and especially during large downward ramps, where RMSE decreases from 296.10 to 210.10 W/m² for drops exceeding 250 W/m². Whereas a fixed equal-average blend of the two heads reaches 129.62 W/m² RMSE, which does not improve over the Trajectory Head. In contrast, the learned adaptive blend reduces RMSE to 126.69 W/m² and MAE to 75.30 W/m², improving over fixed averaging by 2.93 W/m² RMSE and 2.12 W/m² MAE. This shows that the two decoder outputs are complementary, but that their contribution must be weighted adaptively rather than combined with a fixed average.

The GatePersist variant helps isolate the value of the learned Delta Head. In this variant, the Delta Head is replaced with a hard-coded anchor-persistence branch, the adaptive blend therefore receives less learned visual-irradiance information from the second pathway. As a result, the blend and residual stages provide a limited additional benefit: the Trajectory branch within this ablation achieves 128.23 W/m² RMSE, while the residual-corrected output reaches 128.00 W/m². This suggests that the learned Delta Head does more than provide another forecast; it supplies complementary image-informed structure that makes the adaptive blend and the residual correction more effective, even if that learned branch may also carry some forecast error.

Res Dir (%) is the percentage of samples for which the learned residual correction moves the Base forecast towards the observed final-horizon GHI by increasing an underprediction or decreasing an overprediction. It does not describe the direction of the irradiance ramp itself.

Table 6 – Overall ablation and extreme-ramp analysis on NREL-TSI at +30 min. RMSE, MAE and ΔRMSE are reported in W/m².

| Variant/output | RMSE | MAE | $\lvert\Delta\rvert \geq 250$ | Drop 250 | Res Dir (%) | $\Delta\mathbf{RMSE}_{\mathbf{Rise250}}$ | $\Delta\mathbf{RMSE}_{\mathbf{Drop250}}$ |
|---|---|---|---|---|---|---|---|
| GateFormer | 125.01 | 70.44 | 303.39 | 289.78 | 61.39 | −26.27 | +15.46 |
| No GHI Token | 125.44 | 71.09 | 303.83 | 282.54 | 68.68 | −24.19 | +26.33 |
| No Residual (Base) | 126.69 | 75.30 | 308.03 | 274.33 | − | − | − |
| GatePersist | 128.00 | 72.76 | 314.00 | 290.56 | 58.29 | +8.12 | −8.74 |
| Equal Average | 129.62 | 77.42 | 301.31 | 246.88 | − | − | − |
| Trajectory Head | 129.23 | 79.12 | 319.48 | 296.10 | − | − | − |
| Delta Head | 139.36 | 84.81 | 294.74 | 210.10 | − | − | − |

The GHI-token ablation suggests that residual direction accuracy is not the only determinant of correction quality. Although removing the GHI token increases overall push-direction accuracy from 61.39% to 68.68%, it reduces the positive-ramp residual gain for rises above 250 W/m² from 26.27 to 24.19 W/m² and worsens negative-ramp degradation for drops below 250 W/m² from 15.46 to 26.33 W/m². Despite the residual sign accuracy being less favourable, this indicates that token-level GHI conditioning provides useful irradiance-state information for estimating the magnitude of upward irradiance changes.

Having established the contribution of the principal components, Tables 7–9 examine sensitivity to the ramp-oriented loss configuration, input-history length, and Delta-head target design.

Table 7 – Ramp-oriented loss sensitivity on NREL-TSI.

| Parameter changed | Setting | Overall RMSE | MAE | Rise ≥ 250 | Drop ≤ 250 |
|---|---|---|---|---|---|
| Paper configuration | — | 125.011 | 70.439 | 319.798 | 289.784 |
| Change Coefficient ($\boldsymbol{\lambda_{chg}}$) | 0 | 126.488 | 70.581 | 326.591 | 295.763 |
| | 1 | 126.293 | 71.313 | 331.781 | 282.193 |
| | 4 | 128.152 | 71.896 | 324.477 | 288.723 |
| Event Coefficient ($\boldsymbol{\lambda_{event}}$) | 0 | 127.007 | 72.453 | 342.006 | 287.789 |
| | 0.075 | 128.065 | 73.100 | 324.492 | 276.407 |
| | 0.30 | 127.572 | 71.804 | 328.995 | 282.023 |
| Event Threshold ($\boldsymbol{\tau_{event}}$) | 75 | 126.874 | 71.191 | 325.832 | 289.729 |
| | 300 | 125.938 | 70.425 | 337.610 | 280.582 |
| Event Scale ($\boldsymbol{s_{event}}$) | 125 | 126.722 | 72.135 | 329.607 | 289.951 |
| | 500 | 126.176 | 70.321 | 334.962 | 289.270 |
| Event Cap ($\boldsymbol{c_{event}}$) | 2 | 125.978 | 72.031 | 327.127 | 268.513 |
| | 8 | 126.176 | 70.321 | 334.962 | 289.270 |
| Change Scale ($\boldsymbol{s_{chg}}$) | 100 | 127.993 | 72.458 | 315.911 | 304.525 |
| | 400 | 126.022 | 70.721 | 332.173 | 272.203 |
| Change Cap ($\boldsymbol{c_{chg}}$) | 1.5 | 126.136 | 71.470 | 340.070 | 282.183 |
| | 6 | 128.007 | 72.881 | 324.684 | 294.415 |

The paper configuration provided the best overall RMSE among the tested settings. Removing change-magnitude weighting degraded overall, positive-ramp, and negative-ramp performance. Removing the asymmetric event term also degraded overall and positive-ramp performance but slightly improved negative-ramp RMSE. The observed behaviour therefore indicates a trade-off between aggregate accuracy and direction-specific extreme-ramp performance rather than a uniform degradation caused by the asymmetric loss.

Table 8 – Input-history sensitivity (NREL-TSI +30). Changes are rel. to four-frame configuration, evaluated on the same samples.

| Historical frames | Look-back duration | ΔRMSE | ΔMAE |
|---|---|---|---|
| 1 | 0 min | +3.39 | +1.54 |
| 2 | 10 min | +3.07 | +2.03 |
| 3 | 20 min | +1.49 | +0.94 |
| 4 | 30 min | 0.00 | 0.00 |
| 5 | 40 min | +1.47 | +1.01 |
| 6 | 50 min | +2.34 | +2.02 |

Table 9 – Delta-horizon sensitivity on NREL-TSI at +30 min. Changes are relative to the final-horizon-only Delta configuration.

| Delta configuration | Delta Targets | Delta to Gate | ΔRMSE | ΔMAE |
|---|---|---|---|---|
| GateFormer | $\Delta y_{t+H}$ | $\hat{y}^{\Delta}_{t+H}$ | 0.00 | 0.00 |
| Multi-horizon Delta | $[\Delta y_{t+1}, \Delta y_{t+2}, \Delta y_{t+H}]$ | $\hat{y}^{\Delta}_{t+H}$ | +4.04 | +2.01 |
| Multi-horizon Delta (Outputs to Gate) | $[\Delta y_{t+1}, \Delta y_{t+2}, \Delta y_{t+H}]$ | $[\hat{y}^{\Delta}_{t+1}, \hat{y}^{\Delta}_{t+2}, \hat{y}^{\Delta}_{t+H}]$ | +4.16 | +2.09 |

Four historical frames achieved the lowest error among the tested NREL-TSI configurations and were therefore retained across all datasets. Predicting Delta values at additional horizons increased final-horizon error in this sensitivity analysis, supporting the Delta Head's role as a final-horizon specialist while the Trajectory Head provides the intermediate forecasts.

To determine whether the improvement arises simply from model capacity or conventional forecast combination, Table 10 compares GateFormer with parameter-matched (PM) single-head models and combinations of independently trained Trajectory-PM and Delta-PM forecasts. The Trajectory-PM models predict either the complete trajectory or only the final horizon, while Delta-PM predicts the final-horizon change. The combination controls use equal averaging, a fixed validation-selected weight, ordinary least squares (OLS) stacking, or a sample-dependent post-hoc gate.

Table 10 – Forecast-space controls on NREL-TSI (+30 min).

| Variant | Control | Param ($\mathbf{M^{+}}$) | RMSE | MAE |
|---|---|---|---|---|
| Trajectory-PM | Single Head | 55.692 | 130.00 | 78.15 |
| Trajectory-PM (H=3 only) | Single Head | 55.691 | 129.82 | 77.70 |
| Delta-PM | Single Head | 55.691 | 134.93 | 81.03 |
| Equal Average | Two Model Combination | 111.383 | 127.37 | 73.39 |
| Validation Weighted Blend | Two Model Combination | 111.383 | 127.08 | 73.38 |
| OLS Stacking | Two Model Combination | 111.383 | 126.94 | 73.84 |
| Post-hoc Dynamic Gate | Two Model Combination | 111.383 | 127.20 | 73.48 |
| GateFormer (Base) | Single Model Dual-head | 55.702 | 126.69 | 75.30 |
| GateFormer Final | Base + Residual | 55.702 | 125.01 | 70.44 |

All parameter-matched single-head controls produce higher error than GateFormer Base, while the complete GateFormer also outperforms the independently trained post-hoc combinations. Relative to the stronger final-only single-head control, GateFormer Base reduces RMSE by 3.13 W/m² and GateFormer Final reduces it by 4.81 W/m². GateFormer Final also reduces RMSE by 1.93 W/m² compared with the strongest conventional combination.

### *6.3. Ramp-Regime Diagnostics*

We next evaluate GateFormer across different final-horizon irradiance-change regimes on NREL-TSI. The regimes are defined using the measured change between the latest observed GHI and the final +30 min target. This separates low-change periods from positive and negative ramp events, allowing us to determine whether the residual head improves the adaptive Base forecast uniformly or only under specific irradiance-change conditions. Figure 2 illustrates representative stable, positive-ramp, and negative-ramp sequences, where positive- and negative-ramp days contain the corresponding ramp type but are not mutually exclusive. While Table 11 provides the corresponding regime-wise quantitative assessment.

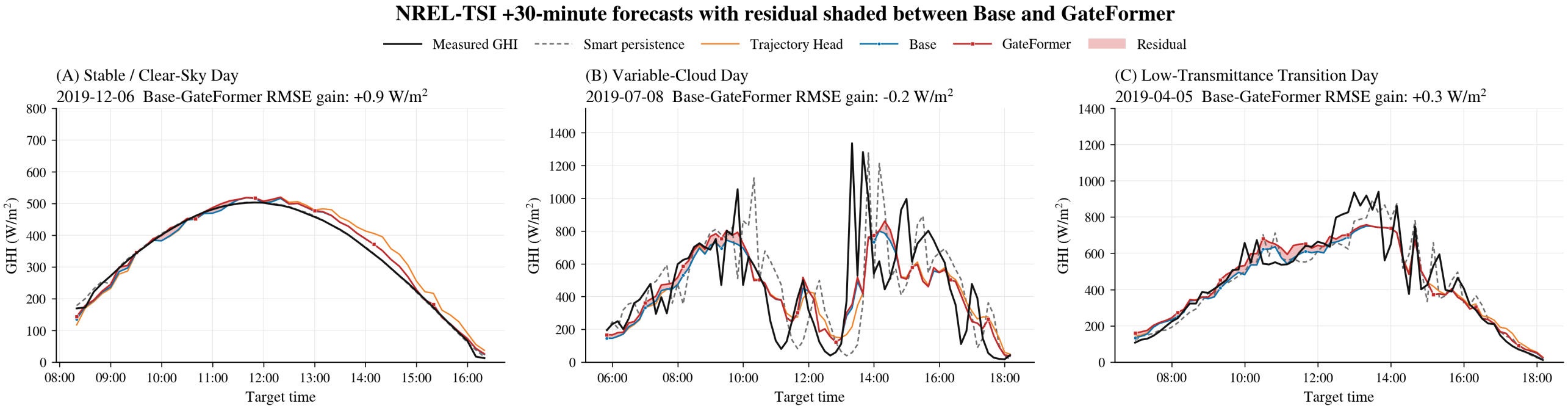

Figure 2 – Representative NREL-TSI +30 min test-set examples across distinct irradiance regimes. Continuous predictions are shown for GHI truth, smart persistence, the Trajectory Head, the Base forecast, and the final GateFormer forecast, together with the predicted residual correction.

Table 11 – Residual diagnostics. ΔRMSE and ΔMAE are calculated as Final minus Base (negative values indicate improvement)

| Regime | n | Base(RMSE) | Final(RMSE) | ΔRMSE | Base(MAE) | Final(MAE) | ΔMAE | Res Dir % |
|---|---|---|---|---|---|---|---|---|
| All | 11188 | 126.69 | 125.01 | −1.68 | 75.30 | 70.44 | −4.86 | 61.39 |
| \|Δ\| < 75 | 5294 | 66.68 | 67.37 | +0.68 | 40.69 | 38.53 | −2.15 | 58.41 |
| Rise 150 | 960 | 274.76 | 253.43 | −21.33 | 224.43 | 199.69 | −24.74 | 85.31 |
| Drop 150 | 1097 | 228.56 | 242.69 | +14.12 | 184.55 | 195.00 | +10.45 | 37.92 |
| \|Δ\| > 150 | 2057 | 251.18 | 247.76 | −3.42 | 203.16 | 197.19 | −5.97 | 60.04 |
| Rise 250 | 497 | 346.07 | 319.80 | −26.27 | 300.39 | 268.22 | −32.17 | 86.12 |
| Drop 250 | 630 | 274.33 | 289.78 | +15.46 | 236.29 | 248.11 | +11.82 | 37.62 |
| \|Δ\| > 250 | 1127 | 308.03 | 303.39 | −4.65 | 264.56 | 256.98 | −7.58 | 59.01 |

Although the proposed framework demonstrates strong performance and overall forecasting accuracy, the residual correction is not equally effective across all regimes: it improves positive ramp events but can degrade negative ramp cases. The residual correction reduces RMSE substantially during positive ramps, with gains of 21.33 W/m² for rises above 150 W/m² and 26.27 W/m² for rises above 250 W/m². However, it degrades negative ramps, where residual direction accuracy is not as strong as its positive ramp counterpart. This identifies negative-ramp correction as a valuable direction for future research.

From a practical perspective, GateFormer adds training cost through its dual decoders and second-stage optimisation but retains a single shared ViT backbone rather than requiring multiple independently trained models. Inference therefore requires one backbone pass followed by comparatively lightweight gating and residual operations. This additional complexity is balanced by consistent accuracy gains across sites and improved positive-ramp forecasting, supporting applications in sub-hourly balancing, storage scheduling, and ramp-rate management using only sky images and measured GHI. Training was completed on a single NVIDIA RTX 3090, although systematic timing comparisons and deployment on lower-resource hardware were not evaluated.

## 7. Conclusion

Operational solar forecasting requires vision models that can convert sky-image sequences into accurate irradiance forecasts. GateFormer addresses this by embedding observed GHI directly into the transformer token stream, producing two complementary forecast decoders, and adaptively blending them before applying a residual correction. On NREL-TSI at +30 min, GateFormer achieved 125.01 W/m² RMSE and 22.97% FS, improving over the SIAT transformer baseline by 9.6% under paired-timestamp evaluation. Across NREL-TSI, Chilbolton, and SIRTA, GateFormer achieved the lowest final-horizon RMSE among the evaluated baselines, with FS ranging from 22.97% to 28.86% relative to smart persistence. Component ablations show that equal averaging does not improve over the stronger Trajectory Head, whereas the learned blend improves over both individual decoders and residual correction provides a further 1.68 W/m² RMSE and 4.86 W/m² MAE reduction after blending.

These results demonstrate that short-term irradiance forecasting benefits from decomposing the prediction into complementary stages: visual-irradiance encoding, trajectory prediction, final-horizon change, adaptive blending, and residual correction. Although negative-ramp correction remains an area for improvement, ramp-regime diagnostics show residual correction improves performance overall and reduces RMSE by up to 26.27 W/m² during positive-ramp events. These gains are relevant to short-term PV balancing, storage scheduling, and ramp-rate management under rapidly changing cloud conditions. Demonstrating that beyond the design of input pathways, feature fusion, and separate system forecast combination, substantial performance gains may also be realised by explicitly structuring how complementary predictions are formed and combined within a shared architecture. By separating representation learning from forecast formation, GateFormer provides a foundation with clear potential for deployment across diverse PV sites using only sky images and measured GHI.

## CRediT Authorship Contribution Statement

**David Hamlyn**: Writing – original draft, Visualisation, Software, Investigation, Conceptualisation. **Sunny Chaudhary:** Writing – review & editing, Software, Supervision, Conceptualisation. **Jonathon Hare:** Writing – review & editing, Supervision, Conceptualisation. **Tasmiat Rahman:** Writing – review & editing, Supervision, Conceptualisation.

## Declaration of Competing Interest

The authors declare that they have no known competing financial interests or personal relationships that could have appeared to influence the work reported in this paper.

## Data Availability Statement

The underlying data and model code are available upon request.

## Acknowledgements

D. Hamlyn acknowledges the support of the Engineering and Physical Sciences Research Council (EPSRC) Doctoral Training Partnership (DTP) Funding grant EP/W524621/1. T. Rahman acknowledges support from EPSRC grant EP/X033333/1. We would like to thank the Chilbolton observatory for providing the data used in this publication.